\documentclass[a4paper,12pt]{article}
\usepackage{amsmath,amssymb,graphicx}
\newcommand{\beq}{\begin{equation}}

\newcommand{\eeq}{\end{equation}}
\newcommand{\myref}[1]{~{(\ref{#1})}}
\newcommand{\mycite}[1]{~{\cite{#1}}}
\newcommand{\eps }{\epsilon}

\def \be  {\begin{equation}}
\def \ee  {\end{equation}}
\def \ba  {\begin{eqnarray}}
\def \ea  {\end{eqnarray}}

\begin{document}
\begin{flushright}

ITEP-TH-07/08\end{flushright}
\begin{center}
{\huge Semiclassical Treatment \\

of Induced Schwinger Processes\\

 at Finite Temperature}\\
\vspace{1.5cm}
 {\large A. K. Monin$^{\dagger\ddagger+}$\\
 {{A. V. Zayakin}}$^{\ddagger * }$\\}
\vspace{1.3cm}
E-mail: \verb"monin@itep.ru, zayakin@itep.ru"\\
\vspace{0.5cm} $^{\dagger}$ M.V. Lomonosov Moscow State
University,\\ 119992, Moscow,
 Russia\\ \vspace{0.2cm} $^{\ddagger}$Institute for
Theoretical and Experimental Physics\\ 117259, Moscow, B.
Cheremushkinskaya 25, Russia\\
\vspace{0.5cm}$^{*}$ Ludwig-Maximilians-Universit\"at M\"unchen,\\ Am Coulombwall 1, D-85748, Garching bei M\"unchen, Germany\\
\vspace{0.5cm} $^+$ School of Physics and Astronomy, University of Minnesota \\
116 Church St. SE, Minneapolis, MN 55455
\end{center}

\vspace{0.5cm}

\abstract{We consider induced pair production in an external field at finite
temperature. One-loop correction to the Green function of a meson is calculated semiclassically within the
framework of saddle-point analysis of Schwinger proper time integrals. This correction appears to be
exponentially small in terms of inverse temperature dependence. Low-temperature limit is shown to be in full
agreement with previously obtained zero-temperature results. The corrections in the low-temperature limits are estimated up to the leading exponential and pre-exponential terms. Comparison is made to earlier calculations of vacuum decay.}\vspace{3cm}
\newpage

\section{Motivation}
Spontaneous processes of particle production in field theory~\cite{Heisenberg:1935qt} (also
known as Schwinger processes) or string/brane production~\cite{Gorsky:2001up}
in string
theory in external fields have long been studied.
Production of $e^+e^-$ pairs by a constant electric field is the
archetypal example for the wide class of these
non-perturbative phenomena. They can generally be characterized by the
essentially non-analytic behaviour of observables in the external
field in the weak field limit, that is, by the presence of $e^{-\frac{1}{E}}$-like
terms. A closely related class of phenomena is known as vacuum decay
processes~\cite{Kobzarev:1974cp,Coleman:1977py}.
One of the similarities between
Schwinger pair production and vacuum decay is that they both can be
described in terms of the semiclassical approximation to the tunneling
problem in quantum mechanics. That is, the leading probability or
another observable is usually organized as $e^{-S_E}$, where $S_E$
is some action on some classical configuration.

The process being spontaneous means the initial state must be a vacuum
state. A generalization of the Schwinger phenomena to the processes in
which a non-zero excitation is contained in the initial state (i.e.
some particle is present) is referred to as induced Schwinger process.
Induced processes of vacuum decay have also been known for quite a
long time~\cite{Affleck:1979px,SelivanovVoloshin}.

Recent progress in understanding induced brane production~\cite{Gorsky:2001up} at zero
temperature has lead us to the following question: how would
finite temperature influence the dynamics of brane production/decay?
This would be of great importance for cosmology.
Brane induced decay is apt to be viewed upon as
false vacuum-decay in a higher-dimensional theory. This task, when
simplified down to field theory level, may be presented as particle
decay in an external field. Recently decay of a magnetic monopole
was calculated by the authors of the present paper within this
string-motivated paradigm, the same was done for a Thirring model meson
decay in 2D theory at zero temperature~\cite{Monin:2006gu}. Here the results  of~\cite{Monin:2006gu} are generalized towards the case of finite temperature.

Studies of spontaneous Schwinger pair production at finite temperatures have a long history. A number of papers have been produced during the last three decades on the spontaneous process of pair production in one-loop approximation. Not claiming to have made a full review in the least part, we cite just a few of them
~\cite{Dittrich:1979ux,Loewe:1991mn,Elmfors:1993wj}.
A modern picture of one-loop thermal results is reflected in~\cite{Gies:1998vt}.
Two-loop results are available as well~\cite{Gies:1999vb}.
On the side of vacuum decay, the finite-temperature effects have been thoroughly studied by~\cite{Garriga:1994ut}  for the spontaneous case, yielding both the exponential and the pre-exponential contributions to the decay rate. On the side of the induced decays, the pre-exponential factor was first calculated in~\cite{Gorsky:2005yq}.

However, no results dealing with temperature corrections to {\it induced} decays are available so far. This was one of the main motivations for writing this paper. While figuring out the simplest one-loop correction to the propagator of a scalar particle due to temperature and external field,  semiclassical treatment is applied which is very close to the world-line techniques by Dunne et al.~\cite{Dunne:2005sx,Dunne:2006st}.
The semiclassical approach to Schwinger processes has been suggested since Popov's papers, see e.g.~\cite{Popov:1973az}.

This article is organized as follows. In Section \ref{tool} a
brief reminder of Green function techniques is given in the finite-temperature
field theory, and the correction to the
meson Green function is calculated, giving a universally  valid  expression (in terms of any regime in  $\beta\equiv \frac{1}{T}$). In Section \ref{betainf}  its asymptotics are studied for  $\beta\to
\infty$, and comments are made on the opposite limit in Section \ref{beta0}. In Section \ref{hier} the problem of leading
and sub-leading asymptotics in semiclassical calculations, as well
as on a relationship between loop resummation and multi-instanton
resummation is briefly discussed. We conclude in Section \ref{conc}.

\section{Schwinger Processes at Finite Temperature}
\subsection{General Techniques\label{tool}}
A simple cubic interaction of a charged scalar $\phi$ and
a neutral $\chi$ scalar in a two-dimensional theory is considered,

\beq\mathcal{L}=\frac{1}{2}|D_\mu \phi|^2-\frac{1}{2} \mu^2|\phi|^2+\frac{1}{2}(\partial_\mu \chi)^2-\frac{1}{2} m^2\chi^2 +\lambda\,\phi\phi^{*}\chi,\eeq where the covariant derivative given as $D_\mu=\partial_\mu+ie A_\mu$.
The masses of the fields being $\mu$ and $m$ are first kept arbitrary, but after proceding to the semiclassical approximation it will be assumed that $\frac{m}{\mu}\ll 1$. This situation is known in vacuum decay terms as ``an almost spherical bubble'' and is used, e.g. in\mycite{Gorsky:2005yq}. The charged field interacts with an external Abelian field $A_\mu$.

One can show \mycite{Schwinger:1951nm}
that in the coordinate representation a free Green function at zero temperature for a particle of field $\chi$ with zero
charge is
\begin{equation}\label{free GF}
G_\chi(x,y)=\frac{1}{(4 \pi)^2}\int_0^{\infty}\frac{d\alpha}{\alpha}\,
\mathrm{e}^{\scriptstyle
im^2\alpha\textstyle-\frac{i(y-x)^2}{4\alpha}},
\end{equation}
where we have omitted the pole prescription $i\varepsilon$.
For the charged particle $\phi$ in the constant external field $A_\mu=(0,E x_0)$ it
becomes
\begin{equation}\label{free GF charged}
G_\phi(x,y)=\frac{1}{(4 \pi)^2}\int_0^{\infty}\frac{\eps d\alpha}{\sinh(\eps
\alpha)}\,\mathrm{e}^{\scriptstyle
im^2\alpha\textstyle-\frac{i\eps}{4}
\scriptstyle(y-x)^2\coth(\eps
\alpha)-\textstyle\frac{i\eps}{4}
\scriptstyle(y_1-x_1)(y_0+x_0)},
\end{equation}
where $\eps=eE$; the field is considered to be far below the Schwinger limit $\frac{m^2}{\eps}\gg 1$. We shall refer to representation\myref{free GF charged} as Schwinger parametrization, and the variable $\alpha$ --- Schwinger parameter.
When the temperature is not equal to zero, the Green function is periodic in the Euclidean
time with a period $\beta\equiv\frac{1}{T}$, hence\mycite{Gies:1998vt} its
generalization is organized as
\begin{equation}\label{GF nonzero T}
\begin{array}{lll}
G_\phi(x,y)&=&\displaystyle\sum_n\int^{\infty}_0
\frac{\eps\,d\alpha}{(4 \pi)^2\sinh(\eps
\alpha)}\,\mathrm{e}^{\scriptstyle
im^2\alpha\textstyle-\frac{i\eps}{4}\scriptstyle
\left[(y_0-x_0+n\beta)^2+(y_1-x_1)^2\right]\coth(\eps
\alpha)}\times\\&\times&\mathrm{e}^{-\textstyle
\frac{i\eps}{4}\scriptstyle(y_1-x_1)(y_0+x_0+n\beta)}.
\end{array}
\end{equation}
The sum over $n$ naturally appears since one should take into
account all the equivalent positions separated by $\beta$ in
Euclidean time, as explained in the cited paper.

\begin{figure}[t]
\begin{center}
\includegraphics[height = 3.5cm, width=5.7cm]{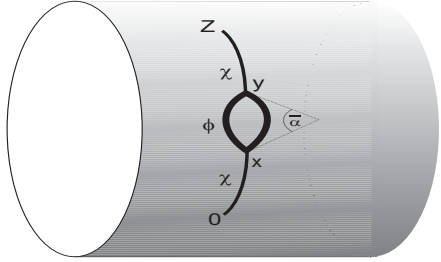}
\caption{One-loop vacuum polarization correction to the propagator of field $\chi$ in an external field at finite temperature. Geometrical meaning of the saddle-point value $\overline{\alpha}$ for the Schwinger parameter $\alpha$ is illustrated; $0,x,y,z$ correspond to the similar variables in \myref{vacpol}; $\phi,\chi$ denote propagator type. } \label{loop}
\end{center}
\end{figure}

Let us consider the one-loop perturbative correction to the Green
function of the uncharged particle due to the cubic interaction mentioned
above
\beq\label{vacpol}
\delta G_\chi(0,z)=\lambda^2\int d^2x d^2y G_\chi(0,x)G_\phi(x,y)G_\phi(y,x)G_\chi(y,z).
\eeq
We think of a one-loop diagram, represented in configuration space in Figure \ref{loop}.
We stress here that our adherence to configuration space representation of vacuum polarization is not an incidental or technical detail of calculation, but is rather of conceptual meaning. Namely, as we have shown in our previous paper~\mycite{Monin:2006gu}, the saddle-point values of $\alpha_i$ correspond directly to the geometric parameters of the classical sub-barrier (Euclidean) trajectory in configuration space. The other reason to keep the position space is the direct relationship of the spatial configuration shown in Figure \ref{loop} to a vacuum bubble with external lines attached to it\mycite{Gorsky:2005yq} describing induced vacuum decay on a compactified manifold.

The correction to Green function of the neutral  $\chi$ field  becomes then
\begin{equation}
\begin{array}{lll}
\displaystyle \delta
G_\chi(0,z)&=&\displaystyle
\frac{\lambda^2\eps^2}{(4\pi)^8}\sum_{p,q,k,n}
\int^{\infty}_0
\frac{d\alpha_1\,d\alpha_2\,d\alpha_3\,d\alpha_4
\,d^2x\,d^2y}{\sinh(\eps \alpha_1)\sinh(\eps
\alpha_2)\,\alpha_3\,\alpha_4} \,\mathrm{e}^{\scriptstyle i\mu^2(\alpha_1+\alpha_2)+i m^2(\alpha_3+\alpha_4)}\times \\
&\times&\mathrm{e}^{\textstyle
-\frac{i(x_0+k\beta)^2}{4\alpha_3}-
\frac{i(z_0-y_0+n\beta)^2}{4
\alpha_4}-\textstyle\frac{ix_1^2}{4\alpha_3}-
\frac{i(z_1-y_1)^2}{4\alpha_4}+\textstyle\frac{i\eps\beta}{2}\scriptstyle(y_1-x_1)(p-q)
}\\
&\times&
\mathrm{e}^{\textstyle-\frac{i\eps}{4}\scriptstyle[(y_0-x_0+p\beta)^2
+(y_1-x_1)^2]\coth (\eps
\alpha_1)\textstyle-\frac{i\eps}{4}\scriptstyle[(y_0-x_0+q\beta)^2
+(y_1-x_1)^2]\coth (\eps \alpha_2)}.
\end{array}
\end{equation}
Note that due to temperature, Lorentz-invariance is explicitly broken in this expression.
We may relate coordinate representation of the Green function correction to
\begin{equation}
\delta G(n,k_1)=\int\delta
G(0;z_0,z_1)\,\mathrm{e}^{-i\omega_nz_0-ik_1z_1}
\,dz_0\,dz_1,
\end{equation}
where $\omega_n=\frac{2\pi n}{\beta}$.
Calculation of this correction corresponds, as usually, to the shift
of the Green's function pole
\begin{equation}
G(n,\vec{k})=\frac{i}{\omega_n^2+k_1^2+m^2+M^2(n,k_1)},
\end{equation}
however, now the shift is not Lorentz-invariant but rather depends on
both $n$ and $\vec{k}$ separately. Whatever complicated expression
for the variation of the Green function one would obtain, physically
relevant information is contained in the pole shift and rescaling of
the wave function. Further on, only the pole shift will be considered.
The mass shift is the value of the $M^2$ calculated at the
point corresponding to the pole. For the zero temperature theory
that means that the shift is determined only by the value of bare
mass (and a scale). Since in this case $M^2(k^2)$ depends only on
$k^2=k_0^2-k_1^2$, one should take $k^2=m^2$. For a theory in a compactified Euclidean space-time it is quite obvious that the shift should
depend on $n$ as $M^2(n,k_1)$, where $n$ and $k_1$ such that
$\omega_n^2+k_1^2+m^2=0$. For the $n$-th Matsubara mode the pole
shift is approximately related to a Green function variation as

\begin{equation}\delta
m^2_n=-\left[\left(m^2+k_1^2+\frac{4\pi^2n^2}{\beta^2}\right)^2\delta
G(k_1,n)\right]_{\omega_n^2+\vec{k}^2+m^2=0}.\label{fourier}
\end{equation}
Moreover the on-shell condition does not have a solution for
arbitrary $k_1^2$ since $\omega_n$ is determined by the discrete
variable $n$. Note again that generically every single mode is
renormalized in its own way. One can easily understand that due to
Lorentz symmetry violation by compactifying the Euclidean time
direction there is no invariant mass anymore. Rather, if some physical quantities related to thermal theory are of interest, the partition function $Z(\beta,\mu,\dots)$  has to be calculated, at some
values of chemical potential $\mu$ and other external potentials. Then some statistically averaged reasonable
quantities have to be found\footnote{The authors are extremely grateful to Professor
H.~M.~Kleinert for a discussion on this point.}, say, $n_i=\frac{\partial
Z}{\partial \mu_i}$, which is the equilibrium
concentration of particles of the $i$-th type.

Therefore,\myref{fourier} is understood merely as a convenient way
of writing down the propagator variation. However, there is a range of parameters
within which it is still possible to preserve the meaning of this quantity as
the mass shift of the particle. This range is the limit of small
temperatures or large $\beta$. In this case on-shell condition can
be solved even for $k_1=0$, since for sufficiently large $\beta$
the value of $\frac{m\beta}{2\pi}$ differs from an integer slightly.
So one can treat the imaginary part of the mass shift for such $n$
and $k_1=0$ as the decay rate of the particle in the external
field with a temperature not equal to zero.

Evaluating elementary integrals, one gets a formal expression for
the mass shift
\begin{equation}
\displaystyle\begin{array}{lll} \displaystyle \delta m^2&=&
\displaystyle\frac{\lambda^2}{\beta\,\eps^{3/2}}\sum\limits_{ r ,
s\in \mathbb{Z} }\delta_{n+r+s}\int\limits ^{+\infty+i0}_0
\frac{d\alpha_1 d\alpha_2}{\sqrt{
\sinh(\alpha_1+\alpha_2)\cosh(\alpha_1-\alpha_2)}} \\
&\times&\displaystyle\mathrm{e}^{\textstyle\frac{4\pi^2
i}{\eps\beta}\frac{(r\tanh
(\alpha_1)-s\tanh(\alpha_2))^2\sinh(2\alpha_1)\sinh(2\alpha_ 2)
}{4\sinh(\alpha_1+\alpha_2)\cosh(\alpha_1-\alpha_2) }} \times \\
&\times& e^{\scriptstyle i\left[r^2\tanh(\alpha_1)+s^2
\tanh(\alpha_2)\textstyle+\frac{\mu^2}{\eps
}\scriptstyle(\alpha_1+\alpha_2)\textstyle+\frac{4\pi^2n^2}{\eps\beta^2}\frac{1}{\coth
(\alpha_1)+\coth(\alpha_2)} \right]}.
\end{array}
\end{equation}

By using the approximate on-shell condition $n\approx
\frac{\mu\beta}{2\pi}$ and Poisson resummation formula
$$
\sum_{n=-\infty}^{\infty}f(n)=\sum_{n=-\infty}^{\infty}
\tilde{f}(2\pi n)
$$
where $$\tilde{f}(k)=\int^\infty_{-\infty}f(t)e^{-ikt}dt,$$
this expression can be written down in two equivalent representations.
Now and further we retain its imaginary part solely, rather then the full (possibly divergent) pole shift in the propagator. These
two representations are

\begin{equation}\label{high-T}
\Gamma=\mathrm{Im}\frac{\lambda^2}{m\beta\,\eps^{3/2}}\sum_{s=-\infty}^{+\infty}\int
\frac{d\alpha_1\,d\alpha_2\,\mathrm{e}^{\textstyle
\frac{i\mu^2}{\eps
}\scriptstyle(\alpha_1+\alpha_2)\textstyle-\frac{im^2}{\eps } \frac
{1}{\coth(\alpha_1)+\coth(\alpha_2)}+{\frac{4i\pi^2A}{ \eps
\beta^2}\scriptstyle(s-s_0)^2}}
}{\sqrt{\sinh(\alpha_1+\alpha_2)\,\cosh(\alpha_1-\alpha_2)}}
\end{equation}
and

\begin{equation}\label{low-T}
\Gamma=\mathrm{Im}\frac{\lambda^2}{m\eps}\sum_{s=-\infty}^{+\infty}\int
\frac{d\alpha_1\,d\alpha_2\,\mathrm{e}^{\textstyle
\frac{i\mu^2}{\eps
}\scriptstyle(\alpha_1+\alpha_2)\textstyle-\frac{im^2}{\eps } \frac
{1}{\coth(\alpha_1)+\coth(\alpha_2)}\textstyle-\frac{i\eps \beta^ 2
s^2}{4A}\scriptstyle-2\pi is s_0}} {\sinh(\alpha_1+\alpha_2)},
\end{equation}
where

$$A=\frac{\sinh(\alpha_1+\alpha_2)}{\cosh(\alpha_1-\alpha_2)},$$
$$s_0=\frac{m\beta}{2\pi}\frac{1}{\tanh(\alpha_2)\coth(\alpha_1)+1}.$$
The sums above can formally be converted to Jacobi theta-functions,
however, that would not be of much use, since the integrals over
Schwinger parameters would then get out of feasibility. On the
contrary, one can do the integrals in Schwinger parameters by
saddle-point method for each term in the sum, provided saddle-point
works at all. Then in principle one could try to do the sum exactly.

When dealing in such way with 1-dimensional saddle-point integrals
$$
\int dz \sum_n \mathrm{e}^{i f_n(z)}
$$
it will be necessary to restrict the domain of applicability of this approximation by imposing the  conditions in saddle-point values $z=\bar{z}_n$ for the $n$th function $f_n$ \mycite{Fedoryuk}:

\begin{eqnarray}
\displaystyle |f_0^{\prime\prime}(\bar{z}_0)|^{3/2}\gg |f_0^{\prime\prime\prime}(\bar{z}_1)|, \label{SaddlePointApplicable0} \\ \notag \\
|f_1^{\prime\prime}(\bar{z}_0)|^{3/2}\gg f_1^{\prime\prime\prime}(\bar{z}_1), \label{SaddlePointApplicable1} \\ \notag  \\
\mathrm{Im}[f_0(\bar{z_0})]\gg \mathrm{Im}[f_1(\bar{z}_1)].\label{NextModeSuppressed}
\end{eqnarray}
Condition\myref{SaddlePointApplicable0} ensures the possibility of doing saddle-point approximation for the zero mode. It checks whether the next-to-leading order terms in the decomposition of $f_0$ may be neglected. Condition\myref{SaddlePointApplicable1} provides the same check for the first mode. To ensure dominance of the zero mode over the first one, we impose\myref{NextModeSuppressed}.

For multi-dimensional expressions the criteria of saddle-point method applicability become more complicated.
Namely, instead of\myref{SaddlePointApplicable0} or\myref{SaddlePointApplicable1} one must require that

\beq\label{ThirdOrderMultivariableCondition}\frac{\partial^3
f}{\partial z_i\partial z_j\partial z_k}P_{il}P_{jm}P_{kn}
\frac{1}{\sqrt{\lambda_l\lambda_m\lambda_n}}\ll 1\eeq \noindent
where $\lambda_i$ are eigenvalues of second derivatives matrix
$\frac{\partial^2 f}{\partial z_i\partial z_j}$, and $P_{ij}$ is
diagonalisation matrix for $\frac{\partial^2
f}{\partial z_i\partial z_j}$, summation implied over $i,j,k,l,m,n$. Further it will be examined whether these conditions are satisfied for a particular saddle-point function under consideration.

\subsection{Limit $\beta\to \infty$\label{betainf}}
The expression\myref{low-T} seems to be an appropriate representation for the $\Gamma$ in the case  $\beta\to \infty$, since even the naive condition of the
saddle point method applicability fails for \myref{high-T}, namely
the factor in the exponent $\frac{1}{\eps\beta^2}$ becomes small. So, let us take \myref{low-T} and make sure that it indeed
corresponds to the low-temperature limit. It is supposed that the
saddle-point is a symmetric point $\alpha_1=\alpha_2=\alpha$. This is possible due to having
particles of identical mass in the internal lines. Since the
saddle-point values of Schwinger proper times, as shown
in~\cite{Monin:2006gu}, correspond to the geometric parameters of
the classical Euclidean loop configuration, only a symmetric
configuration is expected to be the physically relevant solution of
saddle-point equations.

The full ``decay width'' is a sum over Matsubara contributions

$$\Gamma=\sum_{n=-\infty}^{\infty}\Gamma_n.$$
It is expected that the higher the mode, the more suppressed it is. It will be shown below by means of saddle-point integral that this is indeed true
for the zeroth and first modes.
 One can trivially see that the zero Matsubara mode exponential is

$$f_0=\frac{\mu^2}{\eps }2\alpha-\frac{m^2}{\eps }\frac{\tanh(\alpha)}{2},$$
\noindent
identical to that of~\cite{Monin:2006gu}. This function is extremized for $\alpha=\bar{\alpha}_0$ given by
$$\cosh(\bar{\alpha}_0)=\frac{m}{2\mu}.$$
For simplicity the
case of a very light external particle (``almost spherical bubble'') will be considered, that is $\frac{m}{\mu}\ll
1$. This is indeed the case of interest, as for a particle with
$m>2\mu$ the process will become perturbatively allowed. Checking
the conditions of saddle-point
applicability\myref{ThirdOrderMultivariableCondition}, one gets
$$\frac{3}{\sqrt{2} \sqrt{\frac{m \mu }{\epsilon }}}\ll 1.$$ This
condition is satisfied in our setting, due to having a
sufficiently small field $\frac{m^2}{\eps}\gg 1$,
imposed from the very beginning for the applicability of the
saddle-point method (there always must be a significant exponential
suppression). Thus the leading order (zero dual Matsubara mode)
contribution to the sum\myref{low-T} is
$$\Gamma_0\sim\frac{\pi}{\sqrt{\det_{i,j}\partial_i\partial_j f_0}}\,\mathrm{e}^{-f_0(\bar{\alpha})}.$$
In the leading order in both small parameters $\frac{m}{\mu}$ and $\frac{1}{\eps\beta^2}$ one obtains
$$\det_{i,j}\partial_i\partial_j f_0=\frac{4\mu^4}{\eps^2},$$
hence
\begin{equation}
\Gamma_0\sim
\frac{\lambda^2}{m^2\mu}\,\mathrm{e}^{\textstyle-\frac{\pi\mu^2}{\eps}},
\end{equation}
in agreement with~\cite{Monin:2006gu}.
The next-to-leading term is given by the modes with $s=\pm 1$. In
the limit $\eps \beta^2\gg 1, \mu \gg m$ one obtains
$$f_{\pm 1}=\frac{\mu^2}{\eps }2\alpha-\frac{\eps \beta^2}{4\sinh 2\alpha}.$$
That amounts to the saddle-point equations
$$\cosh 2\alpha=\frac{4\mu^2}{\eps^2\beta^2},$$
solved by
$$\bar{\alpha}\approx\frac{\pi i}{4}$$
in the said approximation.
Collecting all the terms, one gets the first correction

\beq\Gamma_{\pm 1}\sim\frac{\lambda^2}{m\eps^2\beta^2}\,
\mathrm{e}^{\textstyle-\frac{\eps\beta^2}{4}}.\eeq
It can be seen that the dependence on a temperature is essentially
non-perturbative.

Upon calculating the second and the third derivatives of $f_{\pm
1}$, one gets the following inequality as the condition for
saddle-point method  applicability by
evaluating\myref{ThirdOrderMultivariableCondition}
\begin{equation}
\frac{16 \mu ^2}{\beta ^3 \epsilon ^{5/2}}\ll 1,
\end{equation}
which can be rewritten as
\begin{equation}
\beta\gg \frac{1}{\mu}\left(\frac{\mu^2}{\eps}\right)^\frac{5}{6}.
\end{equation}
Obviously this condition is fulfilled provided that the temperature
is high enough. The value of the exponential $f_{\pm
1}$ on the saddle point in the leading asymptotics is
$$f_{\pm 1}(\bar{\alpha}_0)=\frac{i}{4} \beta ^2 \epsilon.$$
\noindent In the aforementioned limit the  $s$-th mode will
be suppressed like $e^{-\eps \beta^2 s^2}$. The condition for the
effective suppression roughly is
$$\beta\gg\frac{\mu}{\eps },$$ which is satisfied for $\beta\to \infty$ limit.
Thus one makes sure that the chosen form of the series\myref{low-T}
indeed suites low-temperature region description.

\subsection{Limit $\beta\to 0\,$?\label{beta0}}
Naively, it seems that one can easily perform similar  calculations for the opposite case of extremely high
temperatures. Just using the dual representation for the decay rate
\myref{high-T} and applying the saddle point method one could get the
answer. However, this is not necessarily so for several reasons. In both
representations for $\Gamma$ the pole has already been chosen by setting $k_1^2=0$. It is easy
to see that it is impossible to set $k_1^2=0$ and find such an
integer value $n$ which would satisfy the on-shell condition
$\frac{4\pi^2n^2}{\beta^2}+m^2=0$ for small $\beta$. So, in order to
fulfill the on-shell condition and, thus, to find the proper expression
for the rate in the limit of high temperature, one should keep
$\vec{k}^2$ non-zero, which reflects violation of Lorentz symmetry by introducing a temperature.
It is also worth mentioning that in this case, since one cannot get
rid of the dependence on $n$, it is absolutely necessary to consider the renormalization
for each Matsubara mode separately. The result certainly cannot be interpreted then as
a high temperature decay rate of the initial particle. It is just some correction to the Green function, the physical meaning of which is not well defined. It may be meaningful in a compactified theory rather than in a thermal one. Intuitively one expects that at high temperatures tunneling term $e^{-S_0}$ (instanton contribution to the semiclassical expression) will be dominated over by the over-barrier term $e^{-\beta E} $(``sphaleron'' contribution). However, in the saddle-point analysis performed by us, the ``sphaleron'' term has not appeared in the limit $\beta\to 0$. Thus no final judgement is passed on the applicability of saddle-point method at $\beta\to 0$, but it is doubtful that it can work as directly as it has worked for $\beta\to \infty$.

\subsection{Resummation Hierarchy\label{hier}}
The values of $\bar{\alpha}$ given above are, of course, not unique.
The true solution to the saddle-point equation is a series of roots
like, say, $\bar{\alpha}=\pm i\arccos\frac{m}{2\mu}+2\pi i n, \quad
n\in \mathbb{Z}$. Therefore, an additional resummation to include
all these roots is to be performed in principle. It is not necessary
from a practical point of view, the terms in the series being
suppressed by the factor $e^{-\frac{\mu^2}{\eps }}$, however, this resummation is noted in order to stress the similarity of the saddle-point
configuration, on which the 1-loop integral is essentially
calculated, and the world-line instanton configuration, proposed in
the semiclassical approach by Dunne et al.~\cite{Dunne:2005sx,Dunne:2006st}.

\section{Conclusion\label{conc}}

An example of a calculation of the one-loop thermal
corrections to the propagator of a neutral scalar particle
interacting with a charged one in an external field has been given. We have found the LO thermal correction to decay width in the semiclassical limit for $\beta\to \infty$, in the case of an ``almost spherical bubble'', i.e. $\frac{m}{\mu}\ll 1$, far below Schwinger limit $\frac{m^2}{\eps}\gg 1$, up to its preexponential factor
\beq\delta\Gamma \sim\frac{\lambda^2}{m\eps^2\beta^2}\,
\mathrm{e}^{\textstyle-\frac{\eps\beta^2}{4}}.\eeq
 Thus the result
of~\cite{Monin:2006gu} has been generalized towards thermal/compactified backgrounds. Of peculiar interest would be extension of the presented semiclassical techniques to strings and branes in thermal backgrounds.
Our result supports that of Garriga\mycite{Garriga:1994ut}. Namely, the technique of Garriga does not intercept any correction for vacuum decay in two dimensions at low temperatures, whereas we give an estimate of this correction, and point out that due to its rapid decrease it can't have been noticed in the framework of Garriga's method.

\section*{Acknowledgements}

Authors are indebted to A.~S.~Gorsky for suggesting this problem and
for fruitful discussions, to A.~Abrikosov Jr. for reading the manuscript and to E.~T.~Akhmedov for constant attention to
this work. We thank  G.~Dunne, S.~Gavrilov, H.~Kleinert,  H.~Gies, P.~Koroteev, D.~Levkov, A.~Mironov, A.~Yu.~Morozov, J.~Rafelski, V.~A.~Rubakov, C.~Schubert, S.~Slizovsky and M.~B.~Voloshin for discussions. We are grateful to Michael Sassville and Barbara Tiede for language corrections. One of us (A.Z.) would like to thank D.~V.~Shirkov for his friendly advice and support. A.Z.
also thanks friendly colleagues and staff of FU-Berlin, where this
work was being finished. This work is supported in part by by the DFG Cluster of Excellence MAP,
DAAD Forschungskurzstipendium, RFBR Grant 07-01-0526 (A.Z.); RFBR Grant 07-02-00878 and Scientific
School grant NSh-3036.2008.2 (A.M.).

\end{document}